\documentclass[doublespacing]{elsart}

\usepackage{epsfig}

\usepackage{amssymb}

\newcommand{\bge}{\begin{equation}}
\newcommand{\ede}{\end{equation}}
\newcommand{\fg}[1]{Fig.~\ref{#1}}

\begin{document}

\hfill {}

\begin{frontmatter}


%
\title{Laboratory simulation of cometary x rays using a
high-resolution microcalorimeter}
%
\author[1]{P. Beiersdorfer\corauthref{cor1}} 
\author[1]{H. Chen}
\author[2]{K. R. Boyce}
\author[3]{G. V. Brown\thanksref{brown}}
\author[2]{R. L. Kelley} 
\author[2]{C. A. Kilbourne} 
\author[2]{F. S. Porter}
\author[4]{S. M. Kahn}

\address[1]{Lawrence Livermore National Laboratory, Livermore, California
  94550}
\address[2]{NASA/Goddard Space Flight Center, Code 662, Greenbelt, MD 20771, USA}
\address[3]{Department of Physics and Astronomy, The Johns Hopkins University, 3400 N. Charles st., Baltimore, MD 21218, USA}
\address[4]{Stanford University, Palo Alto, CA 94305, USA}

%

\corauth[cor1]{Corresponding author. Address: Lawrence Livermore National
Laboratory, L-260, Livermore, CA 94550, USA; email: beiersdorfer@llnl.gov; 
fax: xx-1-925-423-2302}
	
\thanks[brown]{Located at NASA/Goddard Space Flight Center.}

\begin{abstract}

X-ray emission following charge exchange has been studied on the University of California Lawrence Livermore National Laboratory electron beam ion traps EBIT-I and EBIT-II using a high-resolution microcalorimeter. The measured spectra include the K-shell emission from hydrogenlike and heliumlike C, N, O, and Ne needed for simulations of cometary x-ray emission. A comparison of the spectra produced in the interaction of O$^{8+}$ with N$_2$ and CH$_4$ is presented that illustrates the dependence of the observed spectrum on the interaction gas.

\end{abstract}

\end{frontmatter}


\newpage

\section{Introduction}

X-ray production by charge exchange has received attention when
a plausible link was established between cometary x-ray emission and solar wind heavy ions. Fully stripped and hydroglenlike carbon, nitrogen, oxygen, and neon, which are part of the solar wind, were thought to interact with gases in the cometary coma, producing K-shell x rays via the charge exchange mechanism \cite{lisse96,haberli97,wegmann98,schwadron00,cravens02}. 

Recently, high-resolution techniques became available to study the x-ray emission of highly charged ions following charge exchange \cite{beiers03e}. These measurements were able to resolve most x-ray lines, even those from levels with high principal quantum number $n$. Because the measurements were performed on an electron beam ion trap, these measurements also included the emission from the $1s2s ~^3S_1$ level, which could not be detected in fast-ion experiments \cite{greenwood00,greenwood01} because of the slow radiative rate associated with this transition. As a result it is now possible to record complete charge exchange induced x-ray spectra in the laboratory and to use them to simulate the observed emission from comets. 

These new capabilities made it possible to show that cometary x-ray emission can be completely described by charge exchange induced x-ray emission \cite{beiers03e}.
Alternative x-ray production mechanisms, which
ranged from lower-hybrid wave heating, scattering of solar X rays by nano-sized dust grains to electron fluorescence and conversion of the kinetic energy of dust particles to x-rays \cite{krasnopolsky97b,bingham97,ip97,owens98,bingham00}, were shown to not be needed to simulate the observed cometary x-ray spectra. 

In the following we present measurements of the spectra produced by
O$^{8+}$ interacting with N$_2$ and CH$_4$ as well as by Ne$^{10+}$ interacting with neutral neon recorded at our facility with a high-resolution x-ray microcalorimeter array.

\section{Experiment}

Our measurements are carried out at the electron beam ion trap facility at the University of California Lawrence Livermore National Laboratory (UC-LLNL). This facility has been used for almost a decade for studying of the x-ray emission of highly charged ions following charge exchange \cite{beiers96d}. The early measurements involved ions with charge as high as U$^{91+}$ \cite{schweikhard98,beiers99c,beiers00a,beiers00e}. 

The charge exchange measurements were enabled by utilizing the so-called magnetic trapping mode of the electron beam ion trap \cite{beiers96d,beiers94c,beiers95f}. In this mode, the electron beam is turned off after the ions have been produced. The ions are confined in the radial direction by the longitudinal field of two superconducting magnets in the Helmholtz configuration, and in the axial direction by a potential on the upper and lower drift tubes. The interaction gas is selected by puffing the desired neutral gas into the trap.

More recently we used the UC-LLNL facility to study the emission of highly charged ions found in the solar wind. In a study of the K-shell emission of O$^{7+}$ and Ne$^{9+}$ we showed that the shape of the K-shell emission spectrum depends on the ion-neutral collision energy below about a few keV/amu \cite{beiers01c}. These measurements were made with a windowless high-purity Ge detector and thus could not resolve the individual X-ray transitions. 

In order to perform high-resolution spectral mesaurements of the x-ray emission, we implemented in 2000 a microcalorimeter array detector on our facility \cite{porter00}. The x-ray microcalorimeter spectrometer (XRS) was originally developed for the Astro-E mission \cite{kelley99}. It consists of a 32-channel, 13 mm$^2$ detector array sensitive to x rays with energy between 200 and 12,000 eV with a resolution of 10 eV. The latter represents more than an order of magnitude improvement over the resolution of the Ge detector used earlier. An upgrade to the higher-resolution (6 eV) microcalorimeter from the ASTRO-E2 mission was implemented in October 2003 \cite{porter04}.

The improvement in resolving power is illustated in \fg{f1}, where we show a spectrum of the Ne$^{9+}$ K-shell emission following charge exchange of bare neon ions with atomic neon. For comparison we show both the data previously obtained \cite{beiers01c} with the germanium detector and the spectrum recorded with the XRS. The individual x-ray transitions emanating from shells with different principal quantum number $n$ are now clearly resolved. A slight difference between the two spectra beyond the difference in resolution arises in the relative magnitude of the Rydberg lines from levels with $n \geq 3$. This is presumably due to differences in the collision energy, i.e., the ion temperature, between the two runs. More measurements are needed to confirm this hypothesis.

The figure clearly illustrates the resonant nature
of charge exchange between a neutral with inonzation potential $I$ and
an ion with charge $q$, in which the electron is preferrentially transferred to a level
with principal quantum number $n_c$ given by 
\begin{equation} 
n_c  \approx \sqrt{ \frac{I_0}{I} \frac{q^2} {1+ \frac{q-1}{\sqrt{2q}} } },
\end{equation}
where $I_0$ is the ionization potential of hydrogen \cite{janev85b}. Moreover, the strong emission from levels with principal quantum number $n \geq 3$ is a signature of the low ion-neutral collision energy in our trap (typically less than 20 eV/amu), as discussed in \cite{beiers00a,beiers01c,perez01}.

The details of the emission from high-$n$ levels depend on the ionization potential of the interaction gas, as shown in the above equation.  Energy
conservation arguments predict that electrons are captured into higher
$n$ levels if the ionization potential of the donor electron 
is lower. This has been shown experimentally in \cite{beiers03e} and is illustrated by the new XRS spectra in \fg{f2}. The emission from the highest-$n$ level shifts from $n = 6$ in the interaction
with CH$_4$, which has a 12.6 eV ionization potential, to $n = 5$ in the interaction with N$_2$, which has a 15.6 eV ionization potential.

\section{Conclusion}

ASTRO-E2, which is scheduled to be launched in 2005, employs a 32-channel microcalorimeter array similar to the one we use at our facility. The satellite is expected to return cometary x-ray spectra with resolution equal or better than that of the spectra we presented. The observations made with ASTRO-E2 will further stimulate laboratory measurements of charge exchange produced x-ray emission from highly charged ions and will provide new challenges for atomic theory to explain the observations.

\section{Acknowledgments}
Work by the University of California Lawrence
Livermore National Laboratory was performed under the auspices of the U.S.\
Department of Energy under contract No.\ W-7405-ENG-48 and supported by NASA's Planetary Atmospheres Program under W-19,938.

%

%





\newpage
\section*{Figure Captions}

Fig. 1. X-ray emission of Ne$^{9+}$ measured with high-purity Ge detector (solid trace) and with the x-ray microcalorimeter (dashed trace). The emission is produced in the interaction of Ne$^{10+}$ ions with atomic neon. The collision energy is a few eV/amu.

Fig. 2. K-shell emission of H-like O$^{7+}$ produced by
charge exchange with CH$_4$ (top) and N$_2$ (bottom). Note changes in the ratio of Lyman-$\delta$ emanating from $n=5$
and Lyman-$\epsilon$ from $n=6$.

\newpage
\begin{figure}
\epsfxsize=400.0pt
\centerline{\epsfbox{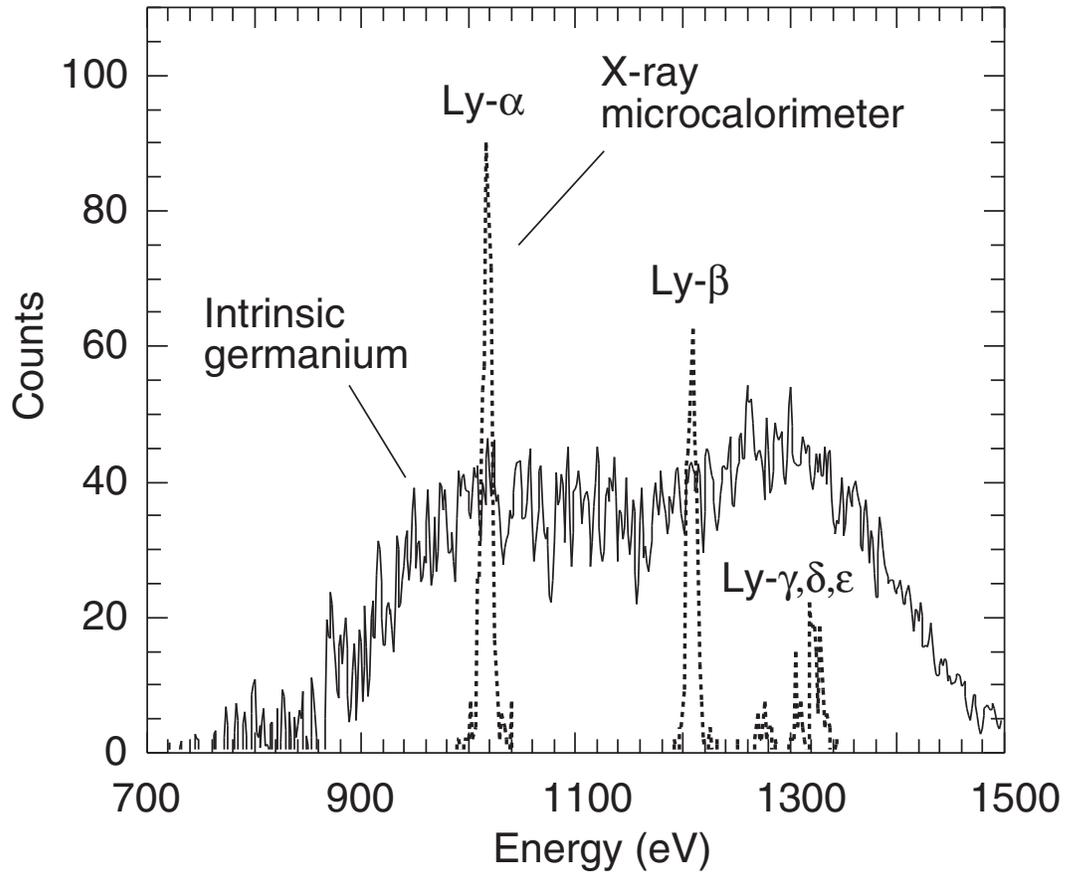}}
\caption{X-ray emission of Ne$^{9+}$ measured with high-purity Ge detector (solid trace) and with the x-ray microcalorimeter (dashed trace). The emission is produced in the interaction of Ne$^{10+}$ ions with atomic neon. The collision energy is a few eV/amu.}
\label{f1}
\end{figure}

\newpage
\begin{figure}
\epsfxsize=400.0pt
\centerline{\epsfbox{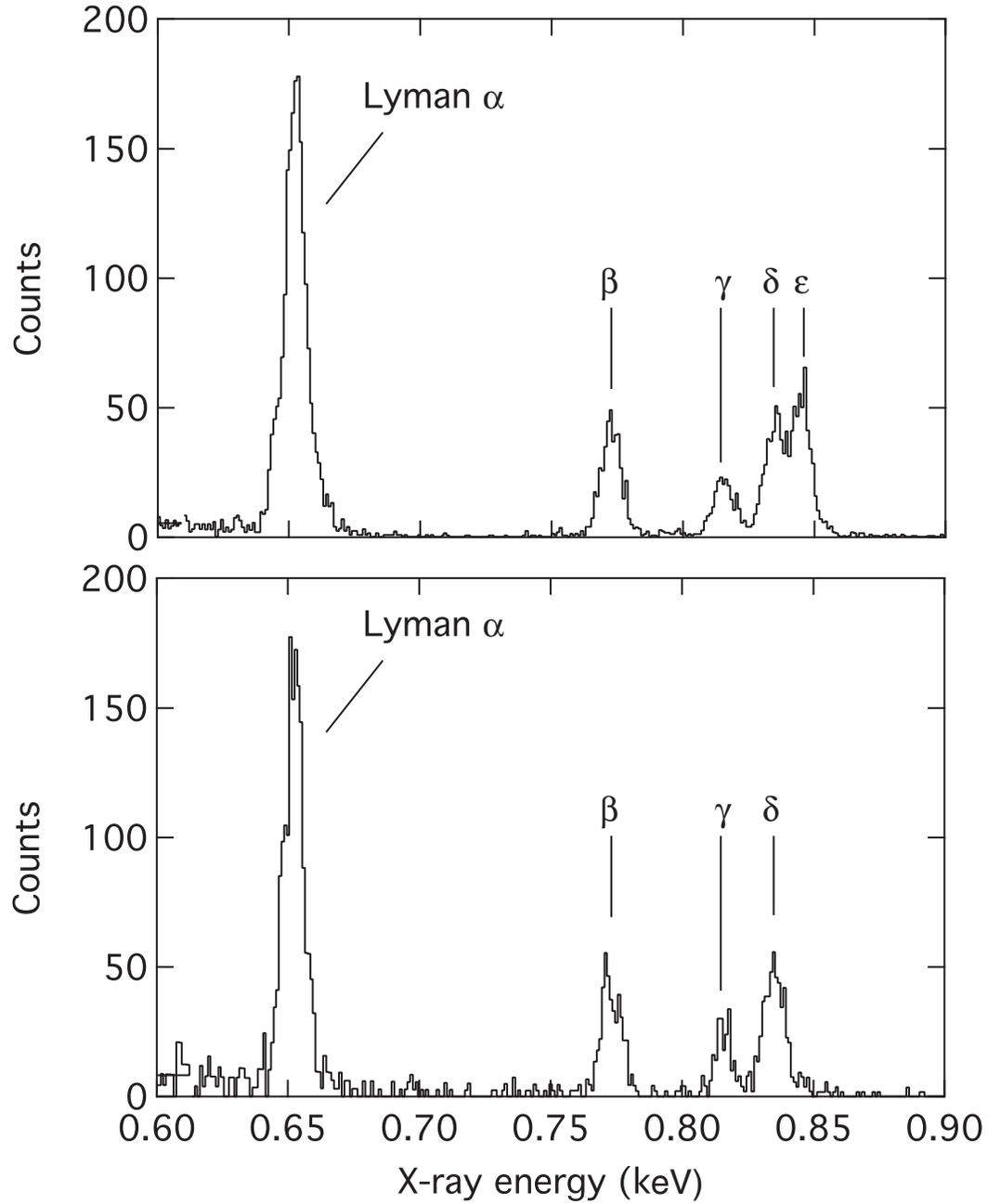}}
\caption{K-shell emission of H-like O$^{7+}$ produced by
charge exchange with CH$_4$ (top) and N$_2$ (bottom). Note changes in the ratio of Lyman-$\delta$ emanating from $n=5$
and Lyman-$\epsilon$ from $n=6$.}
\label{f2}
\end{figure}

\end{document}